\documentclass[12pt]{iopart}

\usepackage{graphicx}
\usepackage[rgb]{xcolor}
\usepackage{iopams}
\usepackage[square,sort,comma,numbers,compress]{natbib}

\usepackage{hyperref}
\hypersetup{colorlinks=true,linkcolor=blue,citecolor=blue,urlcolor=blue}

\begin{document}

\title{Fast manipulation of Bose-Einstein condensates with an atom chip}

\author{R Corgier$^{1,2}$, S Amri$^2$, W Herr$^1$, H Ahlers$^1$, J Rudolph$^1\footnote{Current address: Department of Physics, Stanford University, Stanford, California 94305, USA }$, D Gu\'ery-Odelin$^3$, E M Rasel$^1$, E Charron$^2$, N Gaaloul$^1$}
\address{$^1$ Institut f\"ur Quantenoptik (IQ), Leibniz Universit\"at Hannover, Welfengarten 1, D-30167 Hannover, Germany.}
\address{$^2$ Institut des Sciences Mol\'eculaires d'Orsay (ISMO), CNRS, Univ. Paris-Sud, Universit\'e Paris-Saclay, F-91405, Orsay cedex, France.}
\address{$^3$ Laboratoire de Collisions Agr\'egats R\'eactivit\'e (LCAR), CNRS, IRSAMC, Universit\'e de Toulouse, 118 Route de Narbonne, F-31062 Toulouse, France.}

\eads{\mailto{gaaloul@iqo.uni-hannover.de}}

\begin{abstract}
We present a detailed theoretical analysis of the implementation of shortcut-to-adiabaticity protocols for the fast transport of neutral atoms with atom chips. The objective is to engineer transport ramps with durations not exceeding a few hundred milliseconds to provide metrologically-relevant input states for an atomic sensor. Aided by numerical simulations of the classical and quantum dynamics, we study the behavior of a Bose-Einstein condensate in an atom chip setup with realistic anharmonic trapping. We detail the implementation of fast and controlled transports over large distances of several millimeters, \textit{i.e.} distances 1000 times larger than the size of the atomic cloud. A subsequent optimized release and collimation step demonstrates the capability of our transport method to generate ensembles of quantum gases with expansion speeds in the picokelvin regime. The performance of this procedure is analyzed in terms of collective excitations reflected in residual center of mass and size oscillations of the condensate. We further evaluate the robustness of the protocol against experimental imperfections.
\end{abstract}

\maketitle

%%%%%%%%%%%%%%%%%%%%%%%%%%%%%%%%%%%%%%%%%%%%%%%%%%%%%%%%%%%%%%%%%%%%%%%%
\section{Introduction}
\label{sec:intro}
%%%%%%%%%%%%%%%%%%%%%%%%%%%%%%%%%%%%%%%%%%%%%%%%%%%%%%%%%%%%%%%%%%%%%%%%

Recent proposals for the implementation of fundamental tests of the foundations of physics assume Bose-Einstein condensates (BEC) \citep{RevModPhysCornell2002, RevModPhysKetterle2002} as sources of atom interferometry sensors \citep{articleNaceurVarenna,NJPHartwig2015,ASPALTSCHUL2015,PRLAsenbaum2017}. In this context, atom chip devices have allowed to build transportable BEC machines with high repetition rates, as demonstrated within the Quantus project for instance \citep{ScienceZoest2010,NJPRudolph2015}. The proximity of the atoms to the chip surface is, however, limiting the optical access and the available interferometry time necessary for precision measurements. This justifies the need of well-designed BEC transport protocols in order to perform long-baseline, and thus precise, atom interferometry measurements. 

The controlled transport of atoms is a key ingredient in many experimental platforms dedicated to quantum engineering. Neutral atoms have been transported as thermal atomic clouds \citep{PRLHansel2001, NJPPritchard2006, EPLCouvert2008}, condensates \citep{NatureHansel2001}, or individually \citep{PRBSchrader2001,PRLKuhr2003}, using magnetic or optical traps. Transport of ions with electromagnetic traps has also been achieved recently \citep{PRLBowler2012, PRLWalther2012}. In all those experimental realizations, the transport was performed in 1D. When solving the transport problem, it is tempting to first consider the most trivial solution: the \emph{adiabatic} transport. Yet, besides the fact that the adiabatic solution is far from optimal, it is usually not possible to implement it due to typical experimental constraints. Close to an atom chip surface for example, fluctuations of the chip currents constitute an important source of heating for the atoms, which can lead ultimately to the destruction of the BEC. A nearly adiabatic, and therefore slow, transport is consequently unpractical in most cases. Shortcut-to-adiabaticity (STA) protocols \citep{AMOPTorrontegui2013} were proposed to implement fast, non-adiabatic transport with well defined boundary conditions. Such a reduction of the time overhead can be promising as well for scalable quantum information processing in certain architectures \citep{PhysRevA.80.022303}. On the theoretical side, the protocols that have been proposed relied either on optimal control \citep{PRAPeirce1988, PRAHohenester2007}, counter diabatic driving \citep{Demirplak2003, PhysRevLett.111.100502} or reverse engineering \citep{PRATorrontegui2011}. Besides the transport in harmonic traps, the transport in the presence of anharmonicities \citep{PRAPalmero2013,PRAZhang2015,JPBZhang2016} or the issues related to robustness have been extensively discussed \citep{PRADGO2014}. 

The optimization proposed in this article is found using a reverse engineering method applied to a simplified one-dimensional approximation of the system's classical equations of motion. This solution is then tested numerically in a full three-dimensional quantum calculation using a time-dependent mean-field approach \citep{Pitaevskii1961sovPhys1961, JMPPGPE1963}. Our results are then analyzed in terms of residual center-of-mass and size oscillations of the condensate density distribution in the final trap, at the end of the transport. We then propose to implement a subsequent holding step whose duration is precisely controlled in order to minimize the expansion rate of the BEC in directions where a delta-kick collimation (DKC) procedure \citep{OptLettChu1986, PRLAmmann1997, PRLMuntinga2013, PRLKovachy2015} is not efficient. This DKC step towards the pK regime is necessary for atom interferometry experiments lasting several seconds. The conclusion of this study is that, with the conjugation of ($i$) a controlled transport, ($ii$) a controlled holding time, and ($iii$) a well-designed final DKC step, it is possible to displace BECs by millimeters and to reach expansion speeds in the pK regime. Indeed, the practical implementation we are discussing here leads to an optimal final expansion temperature of 2.2\,pK.

The outline of the paper is as follows: in Sec.\,\ref{sec:model} we describe the architecture of the atom chip and of the magnetic bias field creating the time-dependent potential for the atoms, with strong confinement in two dimensions. We also give the values of currents, bias field, and wire sizes that realize this time-dependent trap. In Sec.\,\ref{sec:theory} we present the theoretical models we are using and their associated numerical implementations, as well as the reverse engineering technique we have adopted. In Sec.\,\ref{sec:results} we give the results of our numerical investigations on the performance of the controlled transport and expansion of the condensate. We also discuss here the robustness of the proposed protocol. Conclusions and prospective views are given in the final section \ref{sec:conclusion}.

%%%%%%%%%%%%%%%%%%%%%%%%%%%%%%%%%%%%%%%%%%%%%%%%%%%%%%%%%%%%%%%%%%%%%%%%
\section{Scheme and atom chip model}
\label{sec:model}
%%%%%%%%%%%%%%%%%%%%%%%%%%%%%%%%%%%%%%%%%%%%%%%%%%%%%%%%%%%%%%%%%%%%%%%%

%%%%%%%%%%%%%%%%%%%
\subsection{Scheme}
%%%%%%%%%%%%%%%%%%%

In this section we introduce the atom chip model and the trapping potential used in the present study. Atom chips designed for the manipulation of neutral atoms are insulating substrates with conducting micro-structures such as metal wires \citep{NatureHansel2001, AcademicPressFolman2002, RevModPhysFortagh2007}. The wire geometry design can easily be adapted for a particular application \citep{AtomChipsReichel2011}. DC wire currents generate inhomogeneous magnetic fields which can be used to trap atoms near the chip surface where high magnetic field gradients produce high trap frequencies and enable fast evaporation. This allows high-flux BEC creation of typically $10^{5}$ atoms/s \citep{NJPRudolph2015}.

\begin{figure*}[t!]
\includegraphics[width=\textwidth]{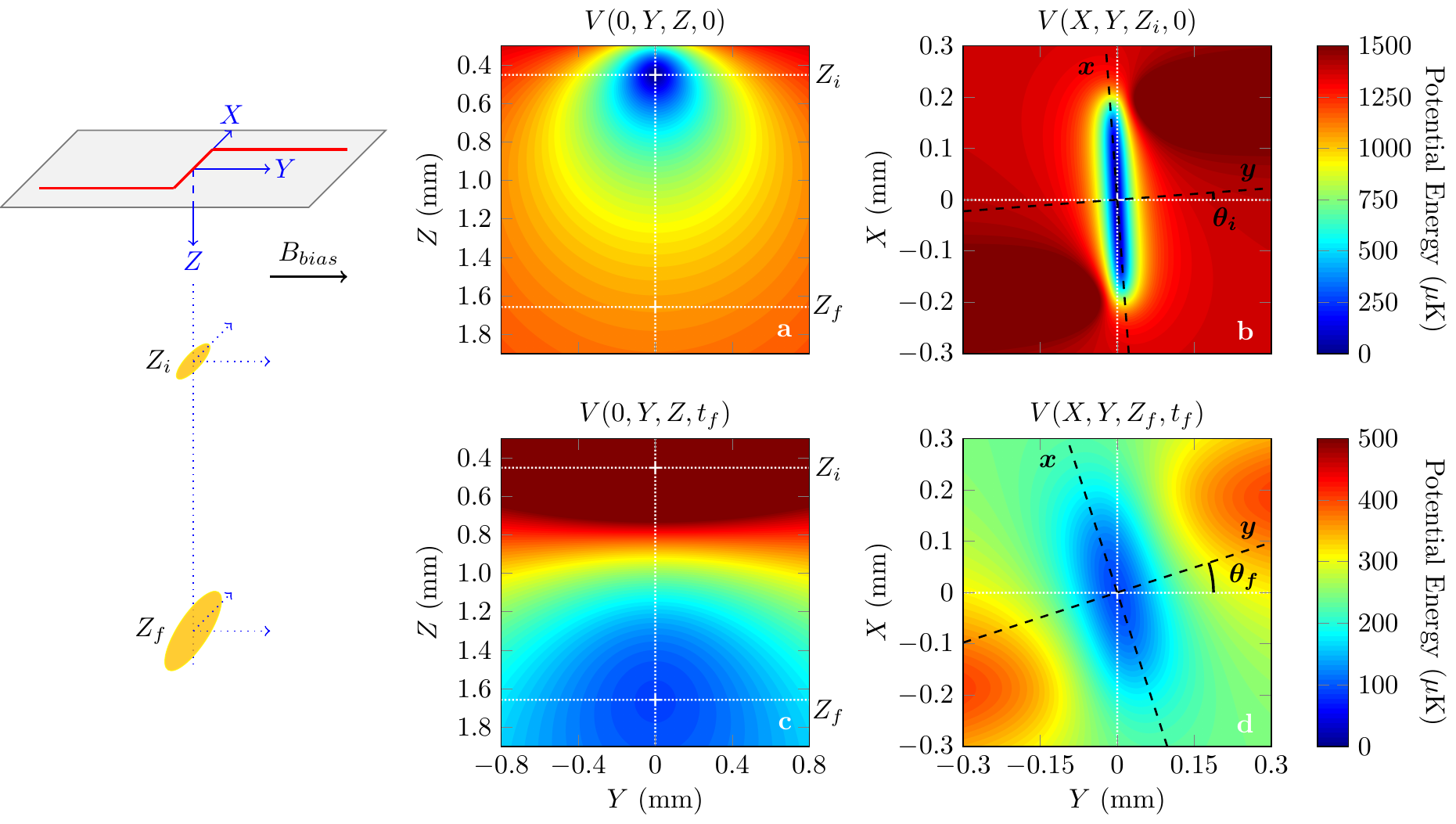}
\caption{\emph{Left panel}: Schematic representation of the chip configuration and of the displacement of the BEC. \emph{Other panels}: (a) and (b) show two cuts of the initial trapping potential $V(\mathbf{R},0)$ in the $(YZ)$ and $(XY)$ planes, respectively. (c) and (d) show similar cuts of the trapping potential $V(\mathbf{R},t_f)$ at the end of the transport procedure corresponding to the time $t=t_f$. The dashed black lines in panels (b) and (d) serve to illustrate the tilt angle $\theta(t)$ of the principal axis $x$ and $y$ of the trap in the $(XY)$ plane. The associated energy color scales are given on the right side, in $\mu$K.}
\label{fig:Model_Z_chip}
\end{figure*}

We consider here the case of a Z-shaped chip configuration \citep{PRLNirrengarten2006}, as shown in Fig.\,\ref{fig:Model_Z_chip}, in the presence of a time-dependent homogeneous magnetic bias field $B_{bias}(t)$. If the bias field varies slowly, the spins of the atoms remain adiabatically aligned with the total magnetic field. In the weak field approximation and in the absence of gravity, the trapping potential can be expressed as
\begin{equation}
V(\mathbf{R},t) = m_F\,g_F\,\mu_B\,B(\mathbf{R},t) ,
\end{equation}
where $\mu_B$ is the Bohr magneton, $g_F$ is the Land\'e factor, $m_F$ is the azimuthal quantum number, and $B(\mathbf{R},t)$ is the total magnetic field. The three-dimensional spatial position is denoted by $\mathbf{R} \equiv (X, Y, Z)$. As shown in Ref. \citep{PRLHansel2001}, a temporal variation of the magnetic field can be used to transport the atoms. Our goal here is to design and test a fast transportation scheme for a realistic setup. We show how the implementation of such a scheme is feasible by specializing our discussion to the hyperfine state $|F=2,m_F=2\rangle$ of the ground 5S$_{1/2}$ state of $^{87}$Rb as a study case. This hyperfine state is a low-field seeking state with $g_F=+1/2$  \citep{dataRb}.

%%%%%%%%%%%%%%%%%%%%%%%
\subsection{Chip model}
%%%%%%%%%%%%%%%%%%%%%%%

The Z-shaped wire is represented schematically on the left side of Fig.\,\ref{fig:Model_Z_chip}. In a first approximation the wires are considered as infinitely thin. The two wires aligned along the $Y$-axis are 16\,mm long. The wire along $X$ measures 4\,mm. They carry a DC current $I_w=5$\,A. The magnetic bias field $B_{bias}(t)$ points along $Y$ and its magnitude varies between $B_{bias}(0)=21.5$\,G (initially) and $B_{bias}(t_f)=4.5$\,G (at the end of the displacement). These parameters are close to those used in the Quantus experiment \citep{NJPRudolph2015}. Slices of the initial trapping potential $V(\mathbf{R},0)$ at time $t=0$ in the $(YZ)$ and $(XY)$ planes are shown in panels (a) and (b) of Fig.\,\ref{fig:Model_Z_chip}, respectively. As shown in the left panel of Fig.\,\ref{fig:Model_Z_chip}, $Z$ denotes the distance to the chip surface. The atoms are initially trapped at a distance $Z_i \approx 0.45$\,mm from the chip surface directly under the origin of the axes. The shape of the trap seen in panel (a) shows a strong confinement in the $Y$ and $Z$ directions with similar trap frequencies $\nu_Y(0)$ and $\nu_Z(0)$. On the contrary, the cigar shape seen in panel (b) reveals that $\nu_X(0) \ll \nu_Y(0) \approx \nu_Z(0)$. The initial trap is thus characterized by a strong two dimensional confinement in the $Y$ and $Z$ directions. The initial potential shows a small tilt angle $\theta(0) = \theta_i \approx 1.53$\,deg in the $(XY)$ plane. In Fig.\,\ref{fig:Model_Z_chip}, the positions of the initial and final potential minima are marked by a white ‘+’ sign. The trapping potential $V(\mathbf{R},t_f)$ obtained at the end of the transport ($t=t_f$) is shown in panels (c) and (d) of Fig.\,\ref{fig:Model_Z_chip}. At this time the minimum of the potential is located at a distance $Z_f \approx 1.65$\,mm from the chip surface and is again centered in the $(XY)$ plane. The BEC transport takes place over a total distance $Z_f-Z_i \approx 1.2$\,mm. This distance is much larger than the typical size of the BEC, of a few $\mu$m. The comparison of panels (Fig.\,\ref{fig:Model_Z_chip}a) and (Fig.\,\ref{fig:Model_Z_chip}c), and of panels (Fig.\,\ref{fig:Model_Z_chip}b) and (Fig.\,\ref{fig:Model_Z_chip}d), shows that during the transport the size of the trap along $Y$ and $Z$ decreases a lot while remaining of the same order of magnitude along $X$. Thus, at $t_f$ the aspect ratio is not as large as initially, and $\nu_X(t_f) < \nu_Y(t_f) \approx \nu_Z(t_f)$. The tilt of the potential has increased to $\theta(t_f) = \theta_f \approx 12.5$\,deg. In order to calculate the three eigenfrequencies of the rotated trap, one can diagonalize the Hessian matrix associated to the potential. This allows to rotate the coordinate system by the tilt angle $\theta(t)$, and to define the new coordinates $\mathbf{r} \equiv (x, y, z)$, with $z=Z$, associated with the three eigen-axes of the trap at any time $t$. The rotated axes $x$ and $y$ are shown as black dotted lines in Figs. (\ref{fig:Model_Z_chip}b) and (\ref{fig:Model_Z_chip}d).

%%%%%%%%%%%%%%%%%%%%%%%%%%%%%%%%%%%%%%%%%%%%%%%%%%%%%%%%%%%%%%%%%%%%%%%%
\section{Theoretical model}
\label{sec:theory}
%%%%%%%%%%%%%%%%%%%%%%%%%%%%%%%%%%%%%%%%%%%%%%%%%%%%%%%%%%%%%%%%%%%%%%%%

In the  harmonic approximation the trapping potential generated by the chip can be written as
\begin{equation}
V(\mathbf{r},t) = \frac{1}{2}m\left[ \omega_x^2(t)x^2 + \omega_y^2(t)y^2 + \omega_z^2(t)(z-z_t)^2 \right],
\label{potential_harmonic}
\end{equation}
where $z_t$ denotes the position of the minimum of the trap along the $z$-axis at time $t$ and $\omega_\alpha(t) = 2\pi\cdot\nu_\alpha(t)$ for $\alpha=x$, $y$ or $z$. For a more precise description of the trap, the lowest order anharmonic term (cubic) along $z$ should be included, yielding the anharmonic potential
\begin{equation}
V_{a}(\mathbf{r},t) = V(\mathbf{r},t) + \frac{1}{3} m \omega_z^2(t) \frac{(z-z_t)^3}{L_3(t)},
\label{potential_cubic}
\end{equation}
where $L_3(t)$ determines the characteristic length associated with this third order anharmonic term. For typical chip geometries as reported in Ref. \citep{NJPRudolph2015}, the cubic term is by far the largest correction to the harmonic order.

\begin{figure*}[ht!]
\center
\includegraphics[height=16.5cm]{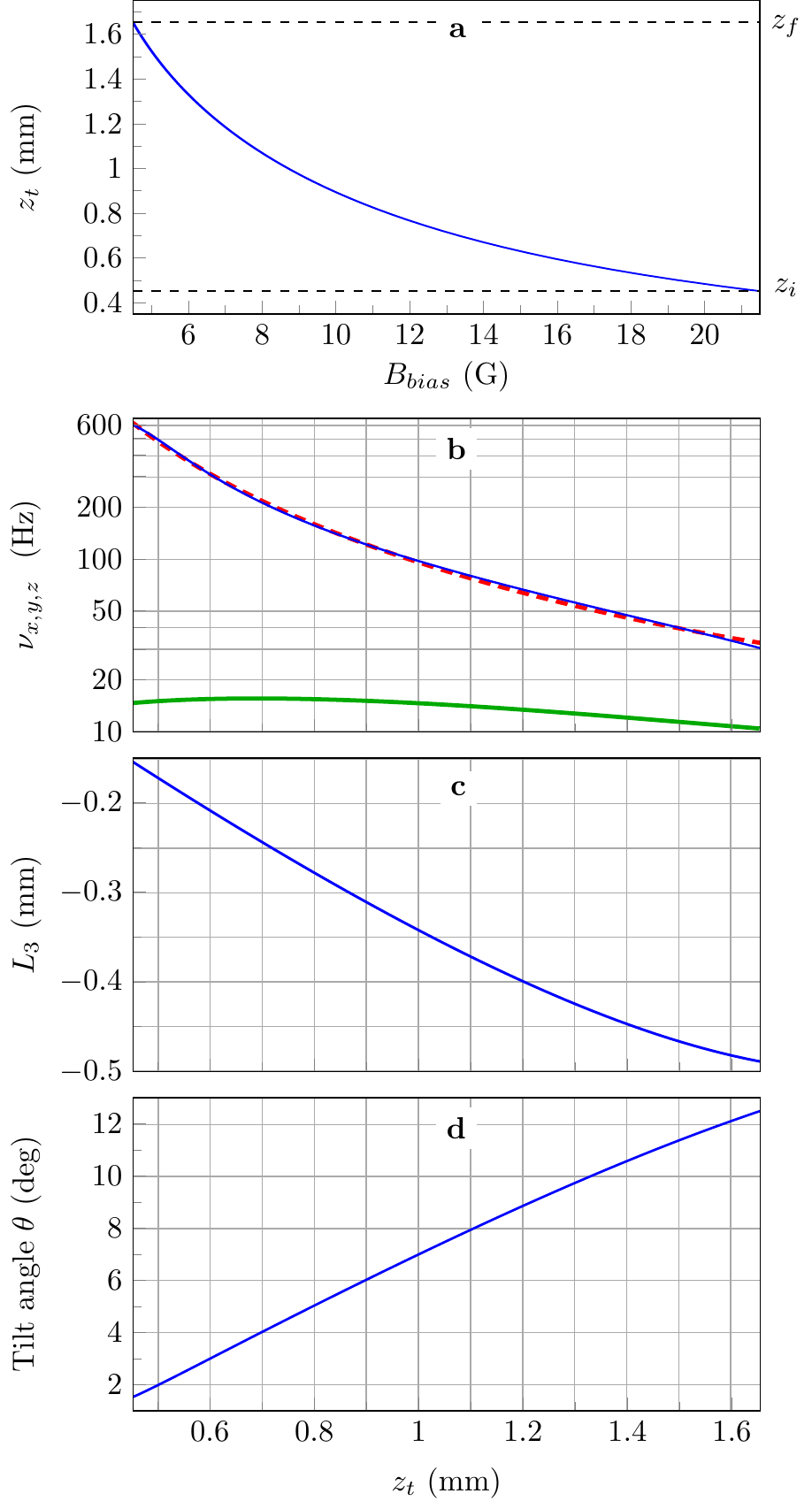}
\caption{(a) Position $z_t$ (in mm) of the minimum of the trap with respect to the chip surface ($z$-direction) as a function of the magnetic bias field $B_{bias}$ (in G). Note that $B_{bias}=21.5$\,G at time $t=0$ and that $B_{bias}=4.5$\,G at the end of the displacement (time $t=t_f$). The two horizontal dashed lines mark the values of the initial and final trap-to-chip distances $z_i$ and $z_f$, respectively. (b) Trapping frequencies $\nu_x$ (thick green line), $\nu_y$ (dashed red line) and $\nu_z$ (thin blue line) in Hz (log scale) as a function of $z_t$ (in mm). (c) Anharmonic coefficient $L_3$ (in mm) as a function of $z_t$ (in mm). (d) Tilt angle $\theta$ (in degrees) as a function of $z_t$ (in mm). For the simulations presented later in the paper, we use accurate analytical fits of the quantities plotted here using second- or (when necessary) third-order Pad\'e approximants\,\citep{site6}.}.
\label{fig:Parameters_Z_chip}
\end{figure*}

Fig.\,\ref{fig:Parameters_Z_chip} shows the different trap parameters used in this study such as the position of the minimum of the trap along the $z$-axis, $z_t$ (Fig.\,\ref{fig:Parameters_Z_chip}a), the trapping frequencies $\nu_x$, $\nu_y$ and $\nu_z$ (Fig.\,\ref{fig:Parameters_Z_chip}b), the parameter $L_3$ (Fig.\,\ref{fig:Parameters_Z_chip}c), and the tilt angle $\theta$ (Fig.\,\ref{fig:Parameters_Z_chip}d). $z_t$ is shown as a function of the experimentally tunable parameter $B_{bias}$ while all other parameters are shown as a function of $z_t$ for the sake of simplicity. In the next section, the theoretical models used to calculate the BEC dynamics are presented.

%%%%%%%%%%%%%%%%%%%%%%%%%
\subsection{BEC dynamics}
%%%%%%%%%%%%%%%%%%%%%%%%%

%%%%%%%%%%%%%%%%%%%%%%%%%%%%%%%%%%%%
\subsubsection{Mean field approach.}
%%%%%%%%%%%%%%%%%%%%%%%%%%%%%%%%%%%%

In the mean-field regime, the evolution of a BEC in a time-dependent potential $V(\mathbf{r},t)$ can be described by the time-dependent macroscopic condensate wave function $\Psi(\mathbf{R},t)$ solution of the Gross-Pitaevskii equation \citep{ Pitaevskii1961sovPhys1961, JMPPGPE1963}
\begin{equation}
i\hbar\,\partial_t \Psi(\mathbf{R},t) =
\left[ -\frac{\hbar^2}{2m}\Delta_{\mathbf{R}} + V_a(\mathbf{R},t) + gN\big|\Psi(\mathbf{R},t)\big|^2 \right] \Psi(\mathbf{R},t),
\label{Eq:GPE}
\end{equation}
where $m$ denotes the atomic mass and $g=4\pi\hbar^2a_s/m$ is the scattering amplitude. $a_s$ is the s-wave scattering length \citep{AtomicPhysicsFoot2005} and $N$ denotes the number of condensed atoms. The non-linear term $gN\big|\Psi(\mathbf{R},t)\big|^2$ describes the mean-field two-body interaction energy \citep{BECPethick2002}. The Gross-Pitaevskii equation is written here in the fixed coordinate system $\mathbf{R} \equiv (X, Y, Z)$, with
\begin{eqnarray}
V_{a}(\mathbf{R},t) & = & \frac{1}{2}m\Bigg[ \,\omega_X^2(t)\,X^2 + \omega_Y^2(t)\,Y^2 + 2\,\omega_{XY}(t)\,XY \nonumber\\
                    &   & \qquad\quad + \omega_Z^2(t)\,(Z-z_t)^2 \bigg( 1 + \frac{2(Z-z_t)}{3L_3(t)} \bigg)\, \Bigg],
\label{potential_cubic_R}
\end{eqnarray}
and
\numparts
\begin{eqnarray}
\omega_X^2(t)  & = & \omega_x^2(t)\cos^2\big[\theta(t)\big] + \omega^2_y(t)\sin^2\big[\theta(t)\big]\\
\omega_Y^2(t)  & = & \omega_x^2(t)\sin^2\big[\theta(t)\big] + \omega^2_y(t)\cos^2\big[\theta(t)\big]\\
\omega_Z^2(t)  & = & \omega^2_z(t)\\
\omega_{XY}(t) & = & \big[\omega_x^2(t)-\omega^2_y(t)\big]\cos\big[\theta(t)\big]\sin\big[\theta(t)\big]\,.
\label{frequency_R}
\end{eqnarray}
\endnumparts

In the present work, the time-dependent Gross-Pitaevskii equation (\ref{Eq:GPE}) describing the BEC dynamics is solved numerically using the second-order split-operator method \citep{JCPFeit1982} and the initial ground state is calculated with the imaginary time propagation technique  \citep{JCPLehtovaara2007}. To efficiently describe the transport over 1.2\,mm while resolving the $\mu$m scale BEC shape on a numerical grid, we use a co-moving frame \citep{PTPTagaki1991, EPJDGaaloul2009, AdvAMOPMEISTER2017}. This transformation eliminates the center-of-mass motion and defines a new time-dependent BEC wave function.
\begin{equation}
\Phi(\mathbf{R'},t) = 
e^{-i\big[\mathbf{K}_a(t)\boldsymbol{\cdot}\mathbf{R'}+\varphi_a(t)\big]}\,
\Psi(\mathbf{R'}+\mathbf{R}_a(t),t)\,,
\label{Eq:FrameTr}
\end{equation}
where $\mathbf{R'} = \mathbf{R} - \mathbf{R}_a(t)$ defines the new time-dependent variable grid, translated by the vector $\mathbf{R}_a(t)$ compared to a fixed laboratory frame, and where
\numparts
\begin{eqnarray}
\hbar\mathbf{K}_a(t) & = & m\dot{\mathbf{R}}_a(t)\,,\\
\hbar\varphi_a(t)    & = & \frac{m}{2}\,\int_0^t |\dot{\mathbf{R}}_a(t')|^2\,dt'\,.
\end{eqnarray}
\endnumparts
Using this transformation, the new time-dependent BEC wave function $\Phi(\mathbf{R'},t)$ verifies the transformed Gross-Pitaevskii equation
\begin{eqnarray}
i\hbar\,\partial_t \Phi(\mathbf{R'},t) & = &
\Big[ -\frac{\hbar^2}{2m}\Delta_{\mathbf{R'}} + V_a\big(\mathbf{R'}+\mathbf{R}_a,t\big) \nonumber\\
& & \qquad
+\,m\,\ddot{\mathbf{R}}_a\boldsymbol{\cdot}\mathbf{R'} + gN\big|\Phi(\mathbf{R'},t)\big|^2 \Big]\;\Phi(\mathbf{R'},t),
\label{Eq:GPE2}
\end{eqnarray}
that we solve numerically in a splitting approach which separates the kinetic energy operator from the potential and interaction energy \citep{JCPFeit1982}. To limit the size of the numerical grid describing the new coordinate $\mathbf{R'} = \mathbf{R} - \mathbf{R}_a(t)$, a good choice for $\mathbf{R}_a(t)$ is to take it as the instantaneous position of the center-of-mass of the BEC \citep{PTPTagaki1991, EPJDGaaloul2009, AdvAMOPMEISTER2017}. Moreover, it is well known that, if the potential remains (to a good approximation) harmonic at all time, the average position of the condensate
\begin{equation}
\mathbf{R}_a(t) \equiv \Big(X_a(t),Y_a(t),Z_a(t)\Big) = \Big\langle \Psi(\mathbf{R},t) \Big| \mathbf{R} \Big| \Psi(\mathbf{R},t) \Big\rangle
\end{equation}
follows Newton's classical equation of motion \citep{Ehrenfest1927}. Since the potential minimum in the $X$- and $Y$-directions remains at the origin, $X_a(t) = Y_a(t) = 0 = x_a(t) = y_a(t)$ applies during the entire propagation. As a consequence the following relations $X'=X$, $Y'=Y$ and $Z'=Z-Z_a(t)$ also hold. Note finally that, along $z$, in the harmonic approximation and in the absence of gravity, the center-of-mass position $Z_a(t) = z_a(t)$ simply follows
\begin{equation}
\ddot{z}_a(t) + \omega_z^2(t) (z_a(t)-z_t) = 0\,.
\label{master_equation_h}
\end{equation} 
In practice, to take into account the eventual influence of anharmonicities we consider that the classical transport trajectory $z_a(t)$ is a solution of the anharmonic equation
\begin{equation}
\ddot{z}_a(t) + \omega_z^2(t)(z_a(t)-z_t)\left(1+\frac{z_a(t)-z_t}{L_3(t)}\right) = 0\,,
\label{master_equation_nh}
\end{equation}
in agreement with Eqs.\,(\ref{potential_cubic}) and\,(\ref{potential_cubic_R}).

As a conclusion, the numerical procedure described in the present section allows to solve the time-dependent Gross-Pitaevskii equation\,(\ref{Eq:GPE2}). It is possible, if necessary, to obtain the solution $\Psi(\mathbf{R},t)$ of the Gross-Pitaevskii equation\,(\ref{Eq:GPE}) in the fixed frame of reference at any time $t$ by simply inverting the relation (\ref{Eq:FrameTr}).

%%%%%%%%%%%%%%%%%%%%%%%%%%%%%
\subsubsection{Scaling laws.}
%%%%%%%%%%%%%%%%%%%%%%%%%%%%%

The introduction of the frame transformation\,(\ref{Eq:FrameTr}) in our numerical approach dramatically improves the computational efficiency of the procedure. In three dimensions, the procedure can remain computationally expensive, depending on the exact propagation time $t_f$ and on the evolution of the trap parameters (position and frequencies). We therefore introduce an alternative, approximate approach, that we compare, in Sec.\,\ref{sec:results}, with the solution of the Gross-Pitaevskii equation.

In the Thomas-Fermi regime of large atom numbers\, \citep{BECPethick2002} and within the harmonic approximation, one can calculate the evolution of the size of the BEC in the rotating frame by solving three coupled differential equations\, \citep{PRLCastin1996, PRLKagan1997}
\begin{equation}
\ddot{\lambda}_\alpha(t)+\omega^2_\alpha(t)\lambda_\alpha(t) =
\frac{\omega^2_\alpha(0)}{\lambda_\alpha(t)\,\lambda_x(t)\lambda_y(t)\lambda_z(t)}\,,
\label{Scaling_law}
\end{equation}
for $\alpha=x$, $y$ and $z$. The scaling factors $\lambda_\alpha(t)$ describe the evolution of the size of the BEC in the three directions, providing that the initial conditions verify $\lambda_\alpha(0)=1$ and $\dot{\lambda}_\alpha(0)=0$. These ``scaling laws'' assume that the BEC keeps its Thomas-Fermi parabolic shape at all time and that the condensate follows adiabatically the rotation of the trap in the ($XY$) plane.

The initial size of the BEC in the rotating frame is then defined by \citep{BECPethick2002}
\begin{equation}
R^{\mathrm{TF}}_\alpha(0)=a_{osc}\,\left(\frac{15 N a_s}{a_{osc}}\right)^{\!\!1/5}\,\frac{\bar{\omega}(0)}{\omega_\alpha(0)}\,,
\end{equation}
where $\bar{\omega}(0) = \big[\,\omega_x(0)\omega_y(0)\omega_z(0)\,\big]^{1/3}$ is the geometric mean of the three oscillator frequencies and $a_{osc} = \big[\,\hbar/m\bar{\omega}(0)\,\big]^{1/2}$ is the characteristic quantum-mechanical length scale of the 3D harmonic oscillator. The characteristic size of the BEC along the three principal axes of the trap $\alpha = x$, $y$ and $z$ is then given at any time $t$ by the relation
\begin{equation}
R^{\mathrm{TF}}_\alpha(t)=\lambda_\alpha(t)\,R^{\mathrm{TF}}_\alpha(0)\,.
\end{equation}
Knowing the parabolic shape of the wave function, these three typical sizes $R^{\mathrm{TF}}_\alpha(t)$ can be related to the three widths $\Delta\alpha(t)$ (\emph{i.e.} the standard deviations in the directions $\alpha=x$, $y$ and $z$) of the BEC wave function, using
\begin{equation}
\Delta\alpha(t) = R^{\mathrm{TF}}_\alpha(t)\,/\sqrt{7}\,.
\label{link_CD-GPE}
\end{equation}
Numerically, we also evaluate the three widths $\Delta x(t)$, $\Delta y(t)$ and $\Delta z(t)$ in the rotating frame from the solution $\Psi(\mathbf{R},t)$ of the time-dependent Gross-Pitaevskii equation in the laboratory frame using
\numparts
\begin{eqnarray}
\Delta x(t)  & = & \Big[ \big(\Delta X\big)^2\cos^2\theta + 2\,\big(\Delta XY\big)\cos\theta\sin\theta
                       + \big(\Delta Y\big)^2\sin^2\theta \Big]^{1/2}\\
\Delta y(t)  & = & \Big[ \big(\Delta X\big)^2\sin^2\theta - 2\,\big(\Delta XY\big)\cos\theta\sin\theta
                       + \big(\Delta Y\big)^2\cos^2\theta \Big]^{1/2}\\
\Delta z(t)  & = & \Delta Z\,,
\label{widths}
\end{eqnarray}
\endnumparts
where
\numparts
\begin{eqnarray}
\Delta X(t)  & = & \big[ \big\langle \Psi \big| X^2 \big| \Psi \big\rangle -
                         \big\langle \Psi \big|  X  \big| \Psi \big\rangle^2 \big]^{1/2} \\
\Delta Y(t)  & = & \big[ \big\langle \Psi \big| Y^2 \big| \Psi \big\rangle -
                         \big\langle \Psi \big|  Y  \big| \Psi \big\rangle^2 \big]^{1/2}\\
\Delta Z(t)  & = & \big[ \big\langle \Psi \big| Z^2 \big| \Psi \big\rangle -
                         \big\langle \Psi \big|  Z  \big| \Psi \big\rangle^2 \big]^{1/2}\\
\Delta XY(t) & = & \big\langle \Psi \big| XY \big| \Psi \big\rangle -
                   \big\langle \Psi \big| X \big| \Psi \big\rangle \big\langle \Psi \big| Y \big| \Psi \big\rangle\,.
\label{widths}
\end{eqnarray}
\endnumparts

%%%%%%%%%%%%%%%%%%%%%%%%%%%%%%%%%%%%%%%%%%%
\subsubsection{Collective excitation modes.}
%%%%%%%%%%%%%%%%%%%%%%%%%%%%%%%%%%%%%%%%%%%

We use the mean-field equation (\ref{Eq:GPE}) together with the scaling approach (\ref{Scaling_law}) to describe the characteristic size excitations of the BEC which arise in the final trap at the end of the transport protocol, due to the fast anisotropic trap decompression over the transport. These excitations can be described as a sum of different collective modes with different amplitudes \citep{PRLCastin1996, PRLStringari1996, PRLMewes1996, PRLKagan1997, PRLDGO1999, RevModPhysDalfovo1999, NJPDubessy2014, JofPRossi2017}. 

\begin{figure*}[ht!]
\includegraphics[width=\textwidth]{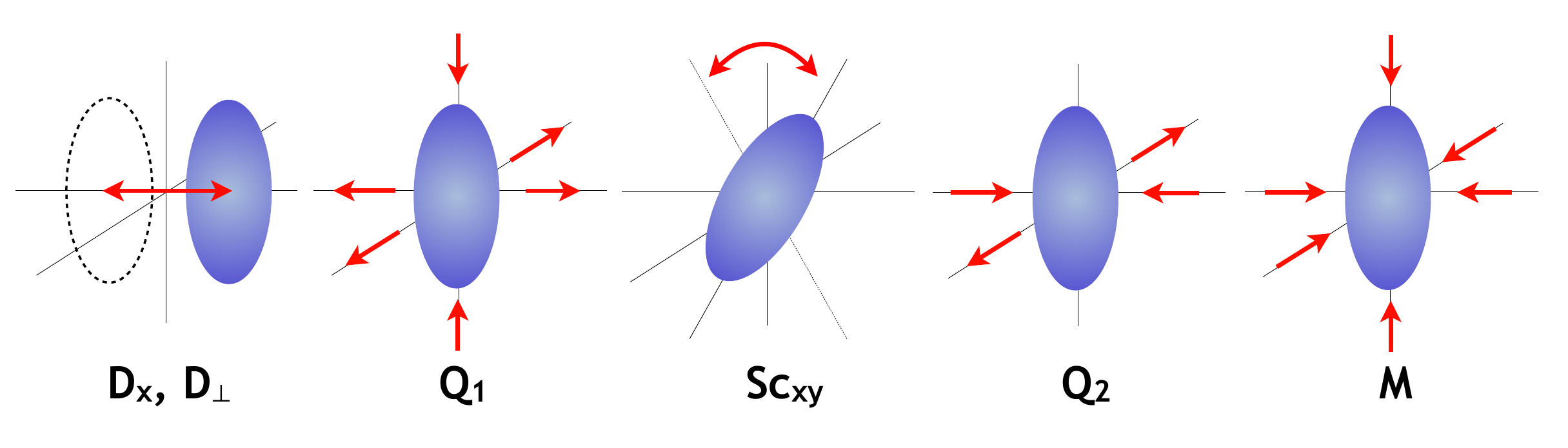}
\caption{Schematic representation of the excitation dynamics of the condensate for the first lowest excitation modes. From left to right: the dipole oscillations $D_x$ and $D_\perp$, the first quadrupole mode $Q_1$, the scissors mode $Sc_{xy}$, the second quadrupole mode $Q_2$, and the monopole mode $M$.}
\label{fig:collective_mode}
\end{figure*}

The first low lying collective excitation modes of a BEC in a cigar shape potential are well known\,\citep{PRLStringari1996}. They can be easily described if we approximate the atom chip trapping potential at time $t_f$ by
\begin{equation}
V(\mathbf{r},t_f) \approx \frac{1}{2} m \omega_\perp^2 \left( \eta^2 x^2 + r_\perp^2 \right)\,,
\label{cylindrical}
\end{equation}
where $r_\perp=\sqrt{y^2+z^2}$ and $\omega_\perp=\omega_y(t_f)\approx\omega_z(t_f)$. The trap aspect ratio is denoted here by $\eta = \omega_x(t_f) / \omega_\perp$. For a low degree of excitation, these modes form a basis of six possible excitations, as depicted schematically in Fig.\,\ref{fig:collective_mode}.

These modes are associated with specific, characteristic frequencies
\numparts
\begin{eqnarray}
\omega_{D_\perp} & = & \omega_\perp \\
\omega_{D_x}     & = & \eta\;\omega_\perp = \omega_x(t_f)\\
\omega_{Q1}      & = & \big[2+3\eta^2/2+\delta/2\big]^{1/2}\;\omega_\perp \\
\omega_{Q2}      & = & \sqrt{2}\;\omega_\perp \\
\omega_{Sc_{xy}} & = & \big[1+\eta^2\big]^{1/2}\;\omega_\perp \\
\omega_{M}       & = & \big[2+3\eta^2/2-\delta/2\big]^{1/2}\;\omega_\perp
\label{law_lying_mode}
\end{eqnarray}
\endnumparts
where $\delta = [9 \eta^4-16\eta^2+16]^{1/2}$. The dipole modes $D_\perp$ and $D_x$ show a classical oscillation of the center of mass of the condensate at the trap frequencies $\omega_{D_\perp} = \omega_\perp$ and $\omega_{D_x} = \eta\,\omega_\perp = \omega_x(t_f)$, respectively. The first quadrupole mode $Q_1$ shows a simultaneous expansion of the two strong axes, while the weak axis is compressed. In the second quadrupole mode, the weak axis does not oscillate and the size oscillations are only present along the two strong axes. The scissors mode $Sc_{xy}$ shows the effect of the trap rotation about the direction of transport, and the monopole mode $M$, also called breathing mode, shows an alternating compression and expansion of the condensate in the three directions in phase.

%%%%%%%%%%%%%%%%%%%%%%%%%%%%%%%%%%%%%%%%%%
\subsection{Reverse engineering protocols}
%%%%%%%%%%%%%%%%%%%%%%%%%%%%%%%%%%%%%%%%%%

We present here the method of reverse engineering, used to find a perturbation-free transport of the center of mass of the BEC\,\citep{EPLSchaff2011, NJPSchaff2011} within a shortcut-to-adiabaticity (STA) approach. This reverse engineering protocol works as follows: we set the classical trajectory of the atoms, $z_a(t)$, according to fixed boundary conditions, which have to be fulfilled experimentally to ensure an optimal transport, \emph{i.e.} initially and finally the center of mass has to be at rest, at the position of the minimum of the trap. This leads to the following boundary conditions
\numparts
\begin{equation}
\label{eq:za_0}
z_a(0) = z_i \quad ; \quad \dot{z}_a(0) = 0 \quad ; \quad \ddot{z}_a(0) = 0
\end{equation}
and
\begin{equation}
\label{eq:za_f}
z_a(t_f) = z_f \quad ; \quad \dot{z}_a(t_f) = 0 \quad ; \quad \ddot{z}_a(t_f) = 0,
\end{equation}
\endnumparts
where $z_i$ and $z_f$ denote the initial and final positions, respectively. To account for experimental constraints, we also wish the trap to be at rest initially and finally. We therefore impose
\numparts
\begin{equation}
\label{eq:zt_0}
z_t(0) = z_i \quad ; \quad \dot{z}_t(0) = 0 \quad ; \quad \ddot{z}_t(0) = 0,
\end{equation}
and
\begin{equation}
\label{eq:zt_f}
z_t(t_f) = z_f \quad ; \quad \dot{z}_t(t_f) = 0 \quad ; \quad \ddot{z}_t(t_f) = 0.
\end{equation}
\endnumparts
The conditions on the second derivatives of the positions are imposed to enforce smooth magnetic field changes. Inserting these last six constraints in Newton's equations (\ref{master_equation_h}) or (\ref{master_equation_nh}) shows that they are equivalent to the additional four constraints
\numparts
\begin{equation}
\label{eq:za34_0}
z_a^{(3)}(0) = 0 \quad ; \quad z^{(4)}_a(0) = 0
\end{equation}
and
\begin{equation}
\label{eq:za34_f}
z_a^{(3)}(t_f) = 0 \quad ; \quad z^{(4)}_a(t_f) = 0
\end{equation}
\endnumparts
where the exponent $(n)$ denotes the $n^{\mathrm{th}}$ time derivative. These four extra boundary conditions (\ref{eq:za34_0})-(\ref{eq:za34_f}) can be seen as additional robustness constraints against oscillations of the center of mass of the BEC in the final trap.
The simplest polynomial solution to the ten boundary conditions can be
fulfilled with the polynomial function of order nine  
\begin{equation}
z_a(t) = z_i + (z_f-z_i)\,\big[ 126\,u^5 - 420\,u^6 + 540\,u^7 - 315\,u^8 +  70\,u^9\,\big]\,,
\label{eq:za_poly}
\end{equation}
where $u=u(t)$ denotes the rescaled time $t/t_f$. The second derivative of the polynomial function (\ref{eq:za_poly}) presents a sine-like variation due to the presence of an acceleration stage followed by a deceleration step. This suggests a non-trivial Ansatz for $z_a(t)$ in the form
\begin{equation}
z_a(t)=z_i+\left(\frac{z_f-z_i}{12\pi}\right)\!\big[6v-8\sin(v)+\sin(2v)\big]\,,
\label{Eric_sin}
\end{equation}
whose second derivative presents a similar sine-like shape, and where
\begin{equation}
v = v(t) = 2\pi \left(\frac{1+a\,u+b\,u^2}{1+a+b}\right) \frac{t}{t_f},
\label{chirp}
\end{equation}
is a `chirped' function of time. The constants $a$ and $b$ act here as two additional control parameters, making this solution more powerful than the simple polynomial one. These parameters can be optimized to limit the impact of the anharmonic term in Eq.\,(\ref{potential_cubic}) in order to recover a BEC at rest after the transport. Note that, according to Eq.\,(\ref{master_equation_nh}), to limit the anharmonic effects, one has to fulfill the following criterium
\begin{equation}
\chi(t)=\left|\frac{z_a(t)-z_t}{L_3(t)}\right| \ll 1\,, \quad \forall t.
\end{equation}
The elaborate form (\ref{Eric_sin}) of $z_a(t)$ is used in Sec.\,\ref{sec:results} with $a=-1.37$ and $b=0.780$. With such parameters the maximum value reached by $\chi(t)$ during the transport is 0.03 while it reaches 0.09 without any chirp (\emph{i.e.} for $a=b=0$). Once $z_a(t)$ is defined, one can extract the magnetic bias field $B_{bias}(t)$ from Eq.\,(\ref{master_equation_h}) since the trap parameters $\omega_z(t)$ and $z_t$ are related to $B_{bias}(t)$ unambiguously. The technical procedure used to extract these parameters, and thus $B_{bias}(t)$, is described in \,\ref{AnnexA}.

%%%%%%%%%%%%%%%%%%%%%%%%%%%%%%%%%%%%%%%%%%%%%%%%%%%%%%%%%%%%%%%%%%%%%%%%
\section{Results}
\label{sec:results}
%%%%%%%%%%%%%%%%%%%%%%%%%%%%%%%%%%%%%%%%%%%%%%%%%%%%%%%%%%%%%%%%%%%%%%%%

The results of the transport protocol realized with the atom chip arrangement described in the preceding sections are presented for a total displacement duration of $75$\,ms. The consequences of this manipulation are evaluated for the position of the wave packet center, denoted as the ``classical'' degree of freedom, as well as for the size dynamics of the BEC.

\subsection{Control of the BEC position dynamics}

\begin{figure*}[ht!]
\includegraphics[width=10cm]{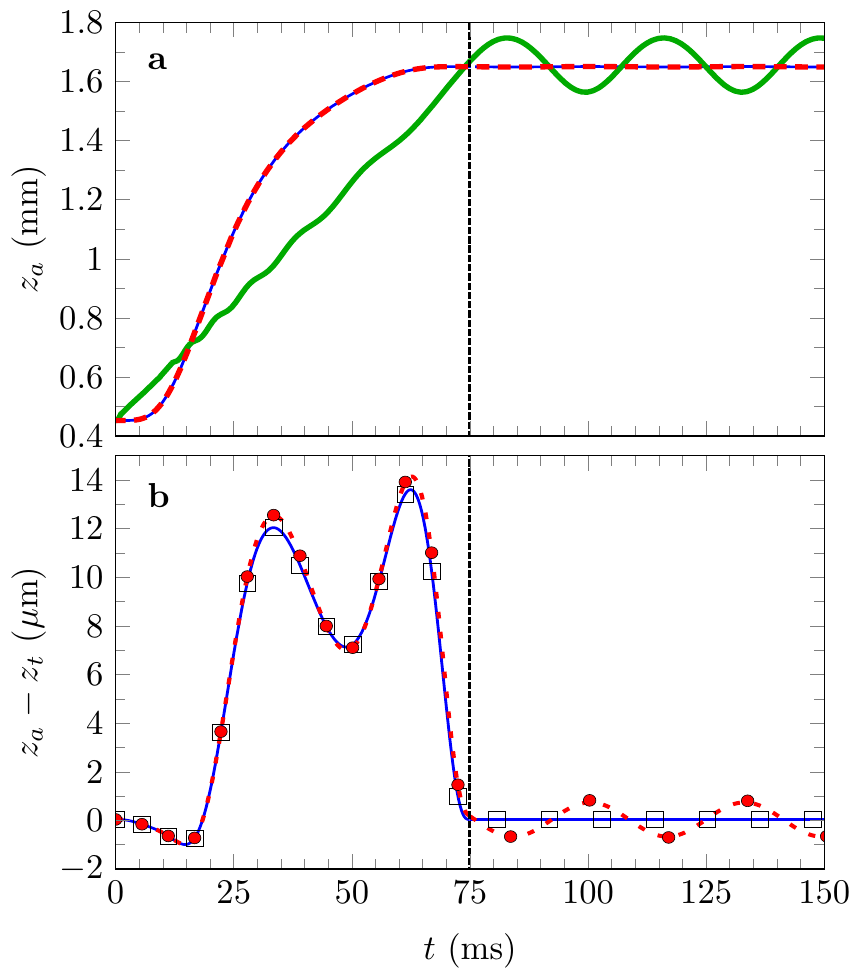}
\centering
\caption{BEC position during and after the STA transport ramp. The vertical dashed line signals the end of the transport time and the beginning of the in-trap oscillations. The upper plot (a) depicts the evolution of the position expectation value $z_a$ of the BEC as a function of time in the case of a linear ramp (solid green), the harmonic trap case (thin blue curve) and the case with a cubic term (dashed red line). The lower graph (b) shows the deviation from the trap position $z_t$ for better visibility of the STA ramp results. The Gross-Pitaevskii solutions are indicated at chosen times by the empty squares (harmonic case) and the plain circles (cubic term included) symbols. In the latter case, the non-adiabatic transport is reflected in residual oscillations of the wave packet in the final trap. Their amplitude is, however, remarkably low ($0.7\,\mu$m).}
\label{fig:COMTransport}
\end{figure*}

In Figure \ref{fig:COMTransport}(a), the atomic cloud position is shown during and after the implementation of the STA protocol in the cases of a chip trap assumed to be harmonic (thin solid blue line) and more realistically including the cubic term of Eq.\,(\ref{potential_cubic}) (dashed red line). In both cases, the classical solution of Newton's equation is indistinguishable from the average position of the wave packet solution of the Gross-Pitaevskii equation. The position of the atomic cloud during the transport is plotted in the left part of the figure. The dashed vertical line signals the end of the displacement and the beginning of a holding period in the final trap. The upper panel (a) of the graph shows the appropriateness of the transport ramp to guide the atoms over more than 1.2\,mm with no noticeable residual center of mass oscillations. To be more convinced that the STA ramp works out thanks to the careful optimization described in the previous section and not because the transport time is long enough to approach the adiabatic limit, we also plot the classical solution for the same displacement time but with a linear ramp (green solid line). The contrast with the optimized solutions is clear, with large residual oscillations with an amplitude of the order of $100\,\mu$m after $t_f=75$\,ms. This clearly shows that the chosen ramp duration is far from the adiabatic time scale, which would trivially bring the atoms at rest in the final trap.

The STA ramp devised in this case allows for the position of the atomic cloud to deviate from the trap position during the transport. This becomes visible in Fig.\,\ref{fig:COMTransport}(b), which shows the offset $[z_a(t)-z_t]$ between the positions of the BEC and the time-dependent trap. In this graph, the Gross-Pitaevskii solutions are indicated at chosen times by empty black squares (harmonic potential) and by plain red circles (cubic term included). For the chosen ramp time, the maximum offset is about $14\,\mu$m. This relatively large offset is responsible for limiting the quality of the transport, as quantified by the amplitude of residual oscillations in the anharmonic case (dashed red line and circles). Indeed, the harmonic solution found for the BEC trajectory by solving Eq.\,(\ref{master_equation_h}) becomes less appropriate the more the atoms explore trap anharmonicities which show up when leaving the trap center. This effect is clearly noticed when comparing the holding trap oscillations in Fig.\,\ref{fig:COMTransport}(b) between the harmonic case (no visible residual oscillations) and the one with a cubic term, which shows an  oscillation amplitude of about 0.7\,$\mu$m around the trap center. It is interesting to note that the chirp introduced in Eq.\,(\ref{chirp}) drastically reduces the residual oscillations of the BEC. Indeed, the oscillation amplitude would reach approximately 6\,$\mu$m with $a=b=0$. Similarly, with no chirp, the same oscillation amplitude of 0.7\,$\mu$m would require a ramp time $t_f > 300$\,ms. The quantum mechanical solution found by computing the average position of the BEC wave function rigorously lies on the Newtonian trajectory in both cases. This was predictable in the harmonic case since it is a consequence of Ehrenfest's theorem applied to our problem. One can also notice that the quantum mechanical solutions are following here the Newtonian trajectory in the anharmonic case. This is because the anharmonic effect is small for such a transport time of 75\,ms.

\begin{figure}[t!]
\centering
\includegraphics[width=9.cm]{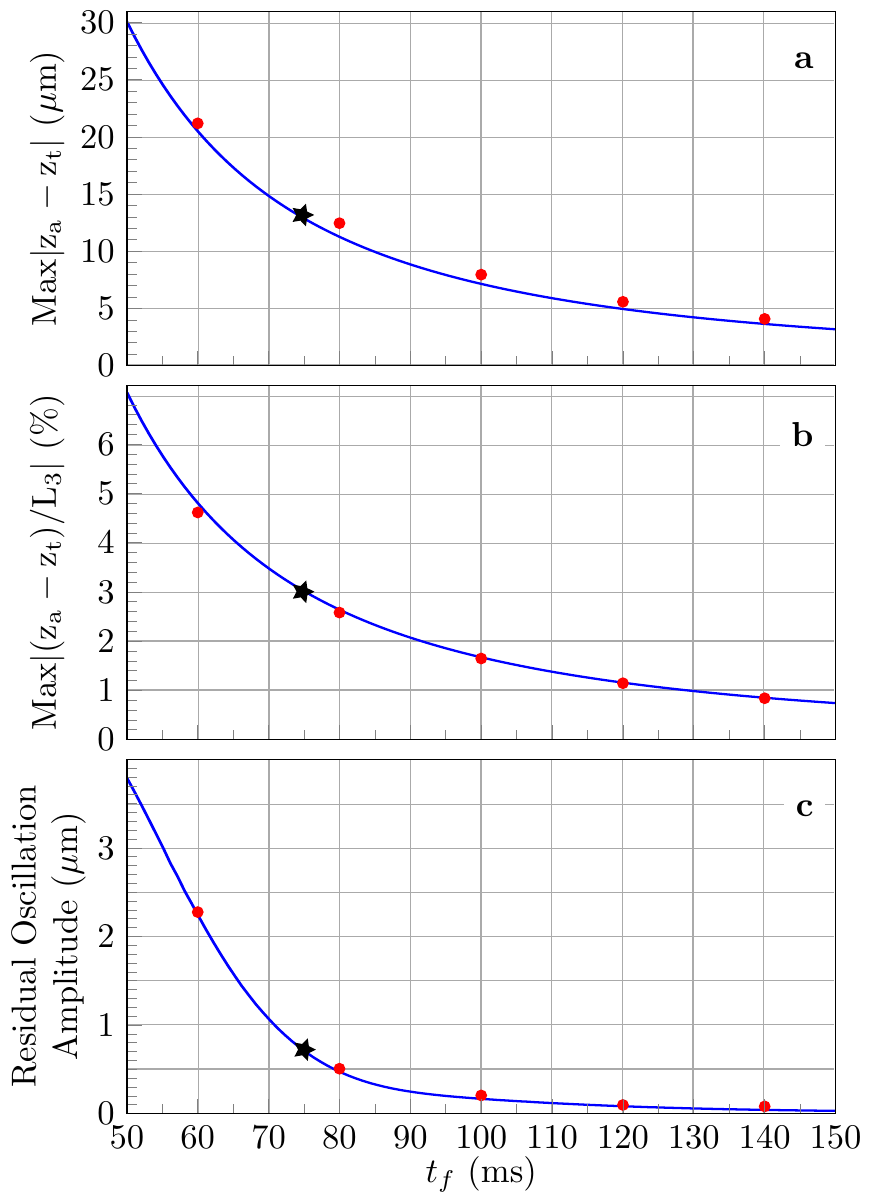}
\caption{Offset to adiabaticity and its impact on the transport as a function of ramp times. The maximum distance reached by the atomic cloud relative to the trap center is plotted in (a), the anharmonicity (cubic potential term) magnitude in percents of the harmonic potential term is depicted in (b) and the consecutive amplitude of the residual oscillations is shown in (c). The longer the transport duration, the more the system tends to the adiabatic limit, the smaller these oscillations caused by the cubic term. The Newtonian trajectories (solid blue curves) agree very well with the full GP solutions (red dots). The black star marks the ramp time $t_f=75$\,ms used in this study.}
\label{fig:Offset_anharmonicity}
\end{figure}

In order to quantitatively assess the magnitude of the anharmonic term during the transport, we plot in Fig.\,\ref{fig:Offset_anharmonicity}(a) the maximum offset to trap center reached by the BEC as a function of the ramp time $t_f$. The solid curve is again corresponding to the Newton's equation solution and the red dots are depicting the Gross-Pitaevskii solution. As expected, short ramps lead to atomic positions departing further from the trap center in both the classical and quantum case since the adiabaticity criterion is less respected. The larger this spatial offset is, the higher the magnitude of the cubic term in Eq.\,(\ref{potential_cubic}) is, the worse the harmonic trap-based reverse engineering for the chip trap trajectory is, and the larger the final residual oscillations are. This is perfectly visible when analyzing Fig.\,\ref{fig:Offset_anharmonicity}(b) giving the magnitude of the cubic term (in percent) relative to the one of the harmonic term. As a consequence, the residual oscillation amplitudes shown in Fig.\,\ref{fig:Offset_anharmonicity}(c) are larger for shorter ramp times, as expected. In all cases, the quantum solution is in a good agreement with the classical one, leading to the conclusion that regarding the position of the wave packets, BECs can here be safely treated as classical point-like particles. As a result, knowing the maximum oscillation amplitude tolerated in an experiment, one can implement our treatment to find the fastest transport ramp.

\subsection{Robustness of the STA protocol}

To assess the practical feasibility of the proposed fast BEC transport, it is necessary to estimate the impact of small experimental imperfections. The present robustness study, therefore, characterizes the residual oscillation amplitude induced by ramp timing errors, denoted here by $\delta t_{\!f}$, and offsets $\delta B_{bias}$ in the time-dependent magnetic bias field applied to drive the chip trap.

Considering the more complete case where cubic potential terms are present, we use Newton's equations (\ref{master_equation_nh}), where $\omega_z$, $z_t$ and $L_3$ are implicit functions of $B_{bias}(t)$. The average position of the condensate can be written as
\begin{equation}
z_a(t) = z_a^0(t) + \epsilon_{z}(t),
\end{equation}
where $z_a^0(t)$ denotes the unperturbed trajectory. A lowest order perturbative treatment applied to Newton's equation (\ref{master_equation_nh}) yields
\begin{equation}
\ddot{\epsilon_z} + \omega_z^2(t)(\epsilon_z-\delta z_t) + \delta\omega_z^2(z_a^0-z_t) + \frac{\omega_z^2(t)}{L_3(t)}(z_a^0-z_t)^2 = 0,
\label{eq:deviation}
\end{equation}
where $\delta z_t$ and $\delta\omega_z$ denote first order perturbations to the trap position and to the trap frequency, respectively. In the following, we solve Eq.\,(\ref{eq:deviation}) for the harmonic (\emph{i.e.} $L_3\rightarrow\infty$) and anharmonic trapping cases.

\begin{figure*}[ht!]
\centering
\includegraphics[width=10cm]{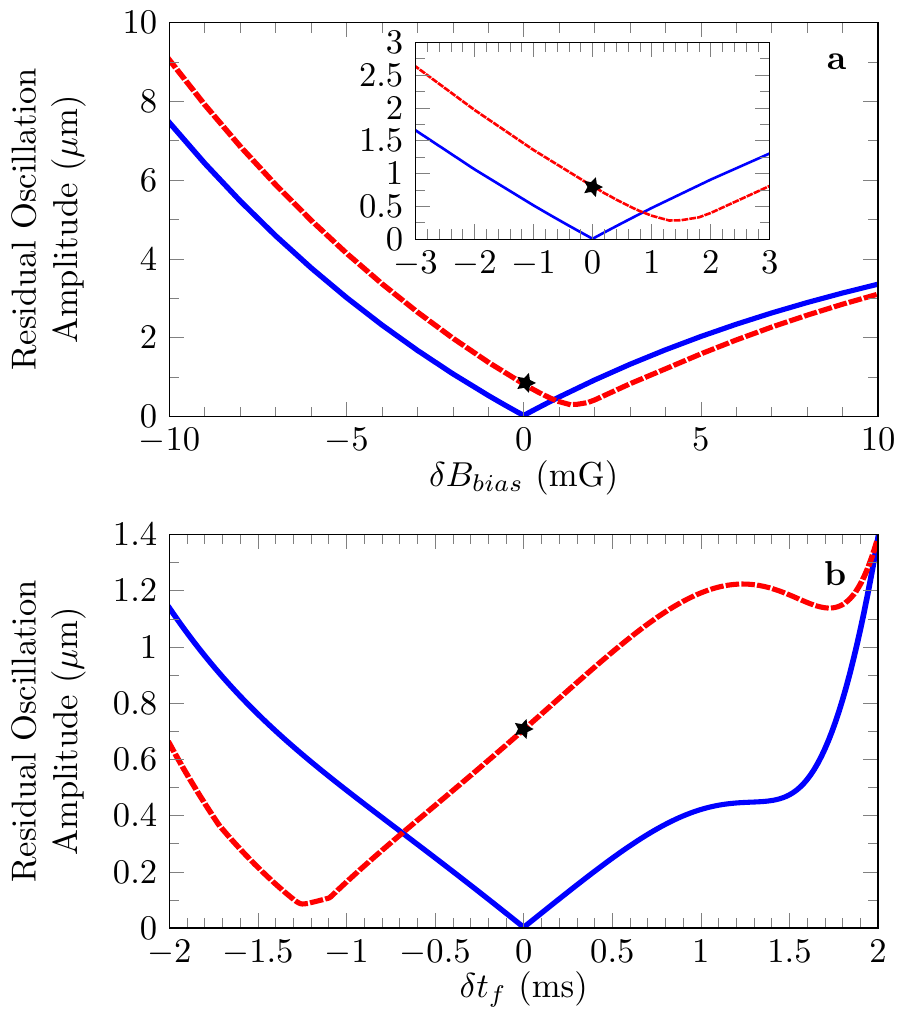}
\caption{Residual oscillation amplitude of the BEC in the final trap as a function of a magnetic field offset (a) or timing errors in the applied ramp (b). Both harmonic (solid blue curve) and anharmonic (dashed red curve) cases are considered. For typical state-of-the-art cold atom experiments, the level of control should be sufficient to ensure errors smaller than 1\,$\mu$m. The black stars mark the results obtained for a ramp time $t_f=75$\,ms taking into account the anharmonicities of the potential in the case where both $t_f$ and $B_{bias}$ are perfectly controlled.}
\label{fig:Robustness}
\end{figure*}

Figure \ref{fig:Robustness} shows the residual oscillation amplitude as a function of the perturbations $\delta B_{bias}$ in panel (a) and $\delta t_f$ in panel (b). This figure confirms the robustness of our transport method. Indeed, $\delta B_{bias} = 1$\,mG of control error in the bias field only leads to an offset of about 0.5\,$\mu$m in the final position of the BEC. Moreover, the same order of infidelity in the final position of the BEC requires ramp timing errors better than 1\,ms, a limit which is easily matched experimentally. Both limits are therefore considered to be safely within state-of-the-art capabilities of standard cold-atom laboratories.

\subsection{Dynamics of the atomic cloud size}

\begin{figure*}[ht!]
\centering
\includegraphics[width=\textwidth]{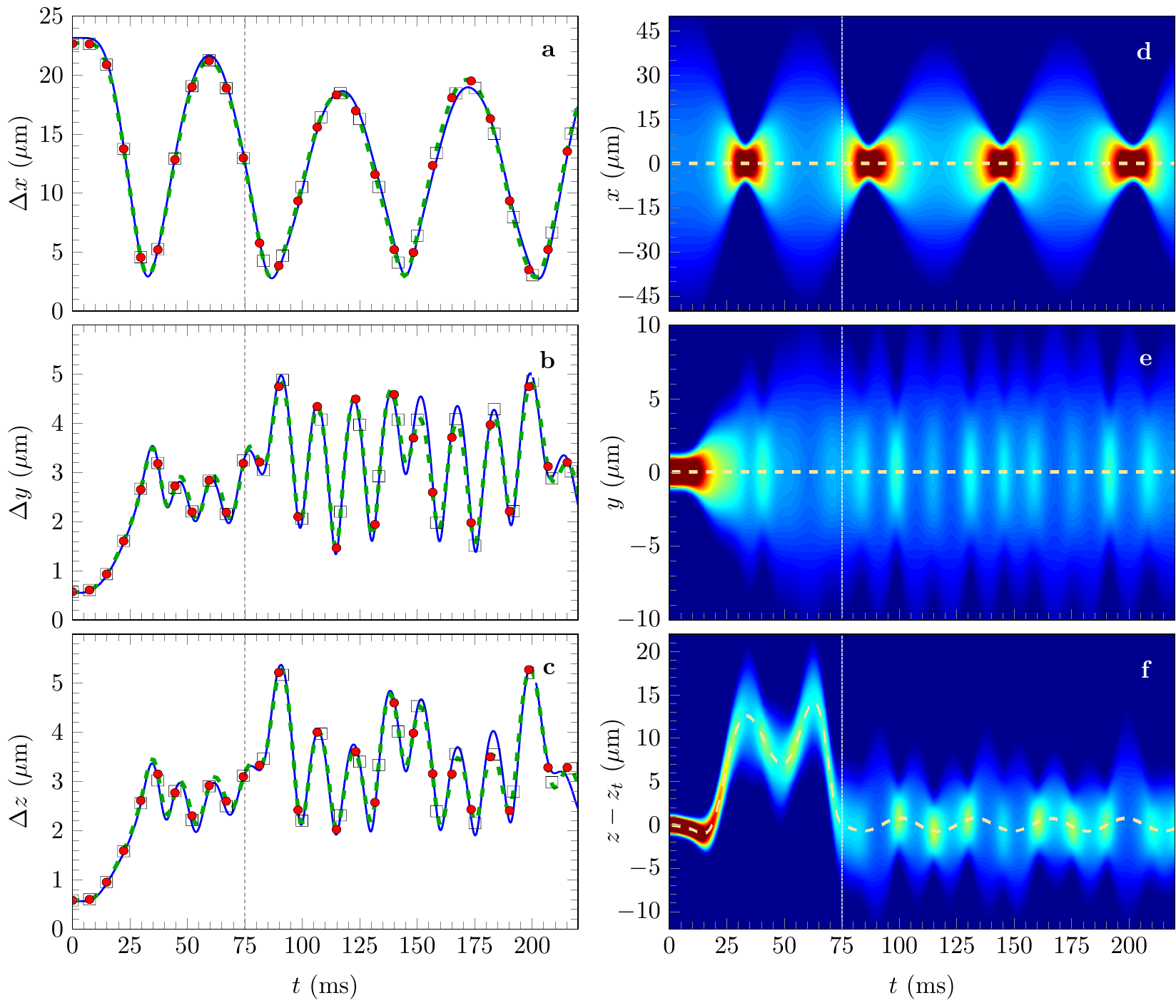}
\caption{Size dynamics of the BEC wave packet. In the left panels (a), (b) and (c), the standard deviations of the spatial density distributions are calculated for the time-dependent condensate wave function for the three principal axes. The solid blue curve is the solution of the scaling approach, the empty black circles are found by solving the Gross-Pitaevskii equation in the harmonic case and the red circles correspond to the more realistic case of the anharmonic trapping potential. The dashed green line is the most complete case including anharmonicities and trap rotation during the transport. The right column shows the averaged probability densities along $x$ (graph d), $y$ (graph e) and $z$ (graph f) calculated by solving the Gross-Pitaevskii equation for the anharmonic case with trap rotation, revealing the collective oscillations connecting the three directions. The dark red regions are associated with density maxima and the dark blue regions correspond to low atomic densities. The last plot (f) is shifted with respect the trap position $z_t$. The dashed orange lines show the expected BEC position in the three directions as a function of time. The vertical dashed lines mark the end of the transport ($t_f=75$\,ms) and the beginning of the holding period.}
\label{fig:Breathing}
\end{figure*} 

In this section, the time-dependent spatial density distribution of the transported BEC is considered. By applying a similar treatment as reported in \citep{EPLSchaff2011, NJPSchaff2011}, it is possible to suppress residual holding trap oscillations of the wave packet center as well as size excitations after the transport. This is achieved, however, at the cost of a long transport time. In this article, we would like to highlight the potential of atom chip-generated STA protocols in the metrological context, \emph{i.e.} with fast enough transport to allow for short duty cycles.

The gallery in Fig.\,\ref{fig:Breathing} shows typical BEC wave packet size oscillations occurring during and after the transport ramp considered in the last section. An adiabatic or long enough transport would bring the BEC to its ground state in the final holding trap, reflected in trivial flat lines starting at $75$\,ms for the three sizes of the left panel. We observe instead a breathing of the wave packet in the three space directions with the largest amplitude occurring in the weak frequency axis $x$. Although the transport is performed in our simulations solely in the $z$ direction, we clearly witness a size oscillation of the atomic wave packet in the two other directions due to the mean-field interactions connecting all spatial directions.

The left panel of Fig.\,\ref{fig:Breathing} illustrates the results of simulations based on the scaling approach (harmonic approximation, solid blue curve) on one hand and on a numerical solution of the Gross-Pitaevskii equation in the harmonic case (black empty squares) on the other. An introduction of the anharmonicities in the Gross-Pitaevskii equation yields the solid red circles and the dashed green line is the most complete case including both, anharmonicities and trap rotation during the transport.

Qualitatively, the four configurations show a similar behavior. The numerical results being similar with and without the cubic term suggests that our trade-off ramp time versus anharmonicities magnitude, previously made for the atomic cloud center, is conclusive regarding BEC size dynamics as well. This is one of the main results of this study since it demonstrates the benign effect of anharmonicities in typical atom chip traps even with fast STA non-adiabatic transports.

The right panel of Fig.\,\ref{fig:Breathing} is a density plot complementing the left part with the density probability distribution during the transport and for 150\,ms of holding time. The quasi-cylindrical symmetry of the trap is reflected in the collective excitation modes observed. Indeed, the strongly trapped directions $y$ and $z$ are subject to in phase size oscillations. The size along the weak axis $x$ is subject to larger-amplitude size oscillations since the trapping frequency is weak along this axis. The excited modes responsible for these oscillations will be identified by the quantitative study of next section.

\begin{figure*}[ht!]
\centering
\includegraphics[width=14cm]{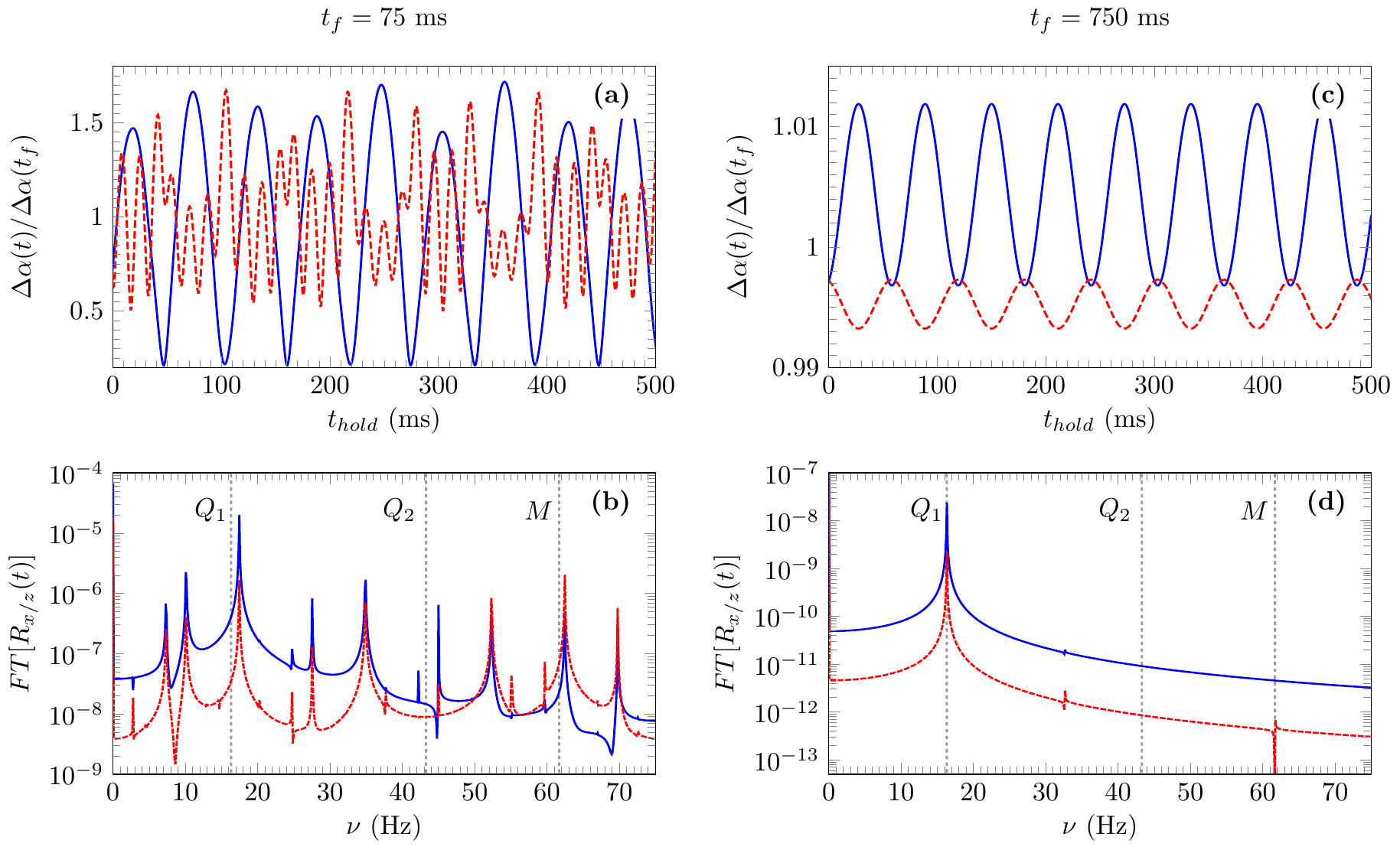}
\caption{\emph{Upper panel}: Size excitation dynamics after the transport of the BEC as a function of the holding time $t_{hold}=(t-t_f)$ in the final trap. Two realizations of the same ramp are considered: a fast transport with $t_f=75$\,ms (left column) and a slow transport with $t_f=750$\,ms (right column). \emph{Lower panel}: Fourier transform of the Thomas-Fermi radius of the BEC in Log scale as a function of the mode frequency $\nu$. For both panels, the solid blue curves are used for the weak axis $x$ and the dashed red ones for the two strong axes $y$ and $z$. The vertical dashed lines in (b) and (d) correspond to the three low-lying excitations modes Q1, Q2 and M calculated following the treatment of \citep{PRLStringari1996}.}
\label{fig:Collective_Modes}
\end{figure*}

\subsection{Collective excitations and optimization of the expansion dynamics}

\subsubsection{Collective excitations in the holding trap}

To gain insight into the impact of the transport speed on the collective excitation of the BEC in the final trap, we plot in Figs.\,\ref{fig:Collective_Modes}(a) and \ref{fig:Collective_Modes}(c) the extracted BEC size oscillations resulting from the ramp of Eq.\,(\ref{Eric_sin}) for a total transport time of $75$\,ms and $750$\,ms, respectively. In order to compare to analytical results, we consider a cylindrical symmetry suggested by Fig.\,\ref{fig:Parameters_Z_chip}(b) where $\nu_y$ is chosen to be strictly equal to $\nu_z$. We plot the sizes normalized to the ones at the end of the transport in the directions $x$ (solid blue line) and $y$ or $z$ (dashed red line) as a function of the holding time $t_{hold}=(t-t_f)$ in the final trap.

In both cases, the final holding time is chosen to be $500$\,ms and one easily notices the complex shape of the residual size oscillations for the fast ramp compared to the slower one, where a simple periodic evolution of the size of the BEC is obtained. This difference occurs due to the rapid variation of the trap aspect ratio in the fast ramp. Indeed, in the transport of Fig.\,\ref{fig:Collective_Modes}(a), the aspect ratio $(\omega_x / \omega_\perp)$ varies by one order of magnitude in 10\,ms only, when a similar variation happens in 100\,ms for the slow transport of Fig.\,\ref{fig:Collective_Modes}(c).

In Fig.\,\ref{fig:Collective_Modes}(b) and (d) we plot the Fourier transforms of the Thomas Fermi radius in Log scale, as a function of the oscillation mode frequency $\nu$ for the two cases mentioned previously of $t_f = 75$\,ms and 750\,ms. These graphs reveal the main collective modes and their harmonics present in the holding trap after the end of the transport. The vertical dashed lines in these plots denote the analytically found collective excitation frequencies according to the treatment described in Sec.\,\ref{sec:theory} and reported in \citep{PRLStringari1996}. This treatment is an approximation in the case of small perturbations. It is clearly not valid for the faster transport reported here. It is, nevertheless, a useful indicator to identify the excitation modes presenting the largest magnitude.

The slow ramp is characterized by the presence of a single quadrupole mode $Q_1$ explaining the simple periodicity of the size oscillation behavior, with the two strong axes in phase and the weak axis out of phase. Note that the oscillation magnitude is, in this case, quite negligible, the size departing only by about $1\%$ from the one at the end of the transport. The fast transport ramp is, however, exciting several collective modes explaining the more complex size oscillation periodicity and the larger amplitude variation to about $\pm 70\%$ change relative to the final transport size in the weak axis $x$.

This analysis is useful on many levels. The predominance of the quadrupole $Q_1$ mode, suggests, for example, the optimization we discuss in the next section. By taking advantage of the symmetry of certain modes, one can also imagine, in further studies, designing a transport protocol forbidding or enhancing them.  

\subsubsection{Optimization of the expansion dynamics}

\begin{figure*}[ht!]
\centering
\includegraphics[width=\textwidth]{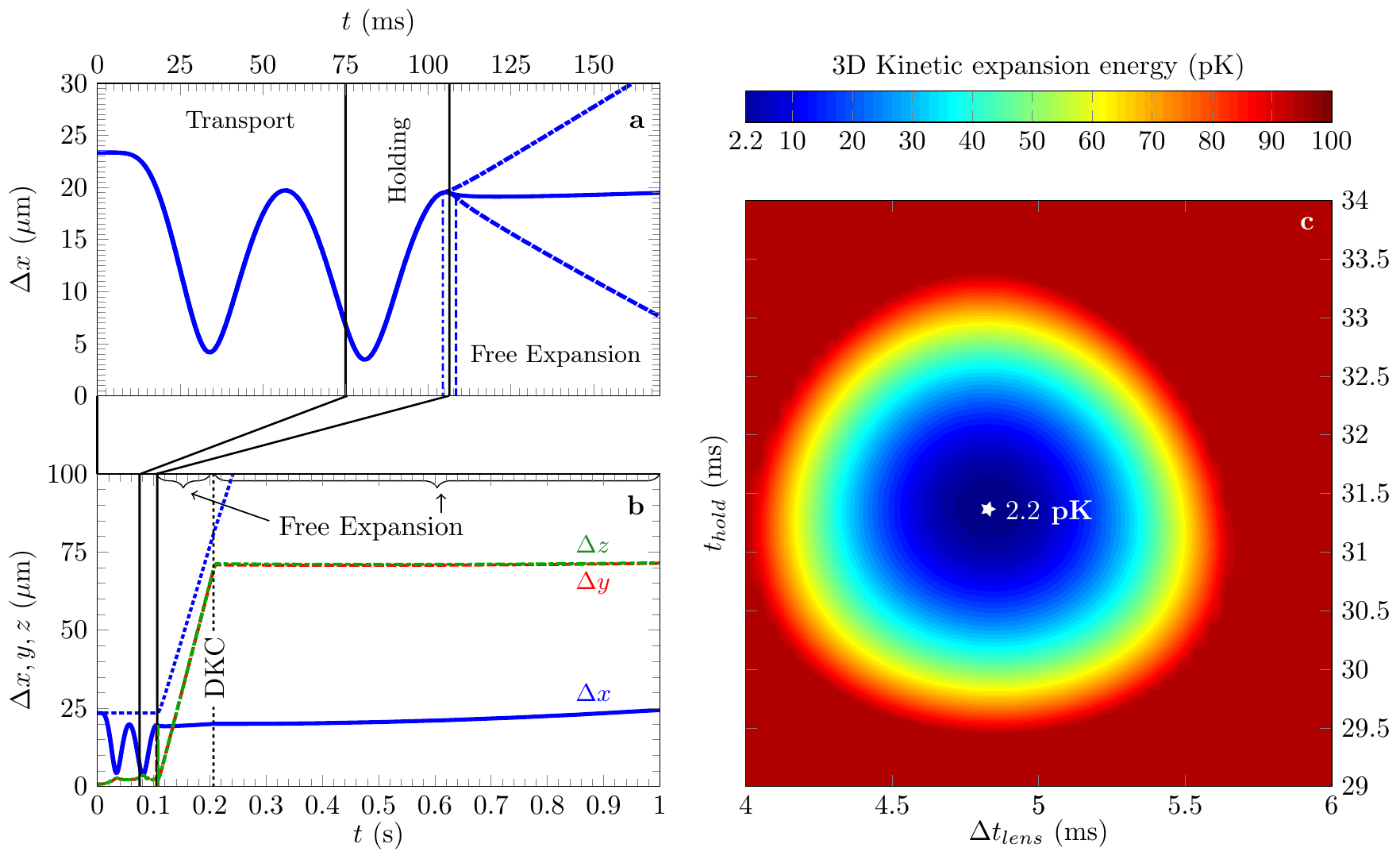}
\caption{Transport, holding, release and magnetic lensing of a BEC to pK expansion velocities. Panel (a): Effect of the release timing from the holding trap in the weak trapping direction $x$. The choices of $29.4$\,ms (dash-dotted blue curve),  $31.4$\,ms (solid blue curve) or $33.4$\,ms (dashed blue curve) illustrate different expansion behaviors (diverging, collimated and focused, respectively). This timing has a little effect on the released size dynamics of the two strong axes $y$ and $z$, not shown here for the sake of clarity. Panel (b): Full sequence with transport, holding, release and delta-kick collimation leading to an average, over the three spatial directions, expansion rate at the pK level. The naturally collimated case of panel (a) is chosen (solid blue line). A free expansion of $100$\,ms is necessary before applying the DKC pulse lasting for $4.84$\,ms in a \{1.7, 7.2, 7.2\} Hz trap. This pulse has a negligible effect on the $x$-axis expansion due to the weak frequency in this direction, but it collimates well the atomic cloud in the $y$ and $z$ directions (dashed red line and dash-dotted green line). The resulting expansion speeds are 22.2\,$\mu$m/s (5.2\,pK), 8.7\,$\mu$m/s (0.8\,pK) and 8.2\,$\mu$m/s (0.7\,pK) in the $x$, $y$ and $z$-directions, respectively. This amounts to a global expansion temperature of $2.2$\,pK only. Without the collective excitations (\emph{i.e.} for an adiabatic transport in the $x$ direction, thin blue dotted curve representing $\Delta x$ as a function of time), the collimation performance is much worse, leading to a global expansion temperature of 555\,pK. Panel (c): Optimal parameters search by scanning the holding and lens durations.  The white star marks the optimal values leading to an expansion temperature of 2.2\,pK shown in (b).}
\label{fig:Expansion}
\end{figure*}

The designed quantum states studied in this article would serve as an input of a precision atom interferometry experiment \citep{AtomInterferometryTino2014}. In such measurements, it is beneficial to work with the slowest cloud expansion possible since this increases the maximum interferometry time available, with an impact on the density threshold for detection and hence, on the sensitivity of the atomic sensor  \citep{AtomInterferometryBerman1997}. Moreover, long free evolution times of several seconds are beneficial for micro-gravity \citep{MGSTRudolph2011, NatrureCommunicationGeiger2011, PRLMuntinga2013, DLRwebsite, CQGAguilera2014, NASAwebsite} and atomic fountain experiments \citep{NJPHartwig2015, PRLZhou2015,PRLAsenbaum2017}. To largely reduce the expansion rate of the atomic samples, the delta-kick collimation (DKC) technique \citep{OptLettChu1986, PRLAmmann1997, PRLMuntinga2013, PRLKovachy2015} is commonly applied. It consists in re-trapping an expanding cloud of atoms for a brief duration in order to align its phase-space density distribution along the position coordinate axis, therefore minimizing its momentum distribution width in preparation for a further expansion. This is in analogy with the collimating effect of a lens in optics and DKC is often referred to as an atomic lens. It is worth noticing that the phase-space density is conserved in such a process which does therefore not achieve a cooling in the sense of reducing the phase-space density. This method was successfully implemented and led to record long observation times of several seconds \citep{PRLKovachy2015, RudolphPHDthesis}.

If the trap is anisotropic, as the quasi-cylindrical case considered in this paper, the lensing effect is different in every direction and would typically be negligible in the weak frequency axis when the two others are well collimated. To overcome this problem, we take advantage of the collective excitations described in the preceding section to release the BEC at a well-defined time, \textit{soon after} a maximum size amplitude of the weak trapping direction such that the subsequent expansion velocity is naturally reduced. This timing is chosen such that the kinetic energy associated with the natural re-compression of the cloud is quickly balanced by the repulsive mean-field interaction energy which naturally leads to an expansion of the cloud in this direction.

To illustrate this optimization, we consider in this section the case of a transport from $z_i \simeq 0.45$\,mm to $z_f \simeq 1.35\,$mm in 75\,ms. The final trap is characterized by the frequencies $\nu_x=12.5$\,Hz, $\nu_y=50$\,Hz and $\nu_z=49.5$\,Hz. This is realized following the reverse engineering technique described in Sec.\,\ref{sec:theory}. The final trap is used to hold the atoms after the end of the transport. The result of this optimization is shown in Fig.\,\ref{fig:Expansion}(a) where the blue curves show the variation of the size of the released BEC in the $x$-direction for three different holding times of $29.4$, $31.4$ and $33.4$\,ms. A natural choice is to consider the switch-off time of $31.4$\,ms leading to a collimated subsequent free expansion. Indeed, a holding time slightly below leads to an immediate fast increase of the condensate size (see the dash-dotted blue line in Fig.\,\ref{fig:Expansion}(a)) while a holding time slightly above leads to a transient compression of the BEC (see the dashed blue line in Fig.\,\ref{fig:Expansion}(a)) soon followed by a very fast expansion.

Following the intermediate and optimal choice $t_{hold}=31.4$\,ms, after $100$\,ms of free expansion the mean field interaction energy is almost entirely in the form of kinetic energy and an atomic lens (DKC pulse) can be applied. It is realized by switching-on a cylindrical trap of frequencies $\nu_x=1.7$\,Hz and $\nu_y=\nu_z=7.2$\,Hz for $\Delta t_{lens}=4.84$\,ms, created with a DC current of intensity $I_w=0.1$\,A and a magnetic bias field of $B_{bias}=0.12$\,G, leaving the trap minimum at $z_f=1.35\,$mm. The collimation effect is dramatic in the $y$ and $z$-directions (red dashed line and green dash-dotted line in Fig.\,\ref{fig:Expansion}(b)). The expansion observed after the application of the DKC pulse corresponds to an average speed in the three spatial directions of about 25.3\,$\mu$m/s, equivalent to a temperature of 2.2\,pK (see \ref{AnnexB} for details). Fig.\,\ref{fig:Expansion}(c) finally shows the robustness of the procedure in case of timing errors for the holding time $t_{hold}$ and for the lens duration $\Delta t_{lens}$. With timing errors as large as 0.5\,ms the expansion temperature remains below 20\,pK. This demonstrates the marginal influence of relatively large timing errors for the 3D collimation effect proposed here.

To illustrate the importance of taking advantage of the collective oscillations, we plot in the same figure the virtual case of an adiabatic transport in the $x$-direction (thin dotted blue line). If one applies a mere adiabatic decompression as suggested by this latter curve, the expansion temperature would be much larger, higher than 550\,pK, even if we consider very well collimated $y$ and $z$-directions, the $x$-direction being hardly affected by the magnetic lens. It is therefore crucial to control the release timing of the BEC in order to implement low-velocity expansions.

%%%%%%%%%%%%%%%%%%%%%%%%%%%%%%%%%%%%%%%%%%%%%%%%%%%%%%%%%%%%%%%%%%%%%%%%
\section{Conclusion and outlook}
\label{sec:conclusion}
%%%%%%%%%%%%%%%%%%%%%%%%%%%%%%%%%%%%%%%%%%%%%%%%%%%%%%%%%%%%%%%%%%%%%%%%

While BEC creation on atom chips was demonstrated with competitive high-flux of $10^5$ BEC atoms/second as a source of metrology-oriented experiments \citep{NJPRudolph2015}, the necessary displacement of the atoms far from the chip surface constrained the use of this technique due to the long times needed to bring atoms at desired positions without detrimental center of mass and size excitations. In this study, we demonstrate a shortcut-to-adiabaticity set of protocols based on reverse engineering that solve the speed issue. This proposal goes beyond existent methods since it includes characteristic mean-field interactions and their coupled effects in the three spatial dimensions even for a 1D transport of a degenerate bosonic gas. To illustrate the appropriateness of our theoretical proposal, we considered the commonly used Z-chip wire geometry combined with a bias homogeneous magnetic field. The study is carried out considering atom-chip-characteristic cubic anharmonic terms in the rotating trapping potentials which manipulate the atoms. Although the STA protocols are inspired by harmonic traps and Newton's equations, the validity of our recipe is supported by solving a scaling approach and mean field equations for interacting BEC ensembles. With the help of analytical and numerical models, we were able to engineer fast atomic transport ramps in few tens of ms and carry a trade-off study between speed and accepted residual excitations at the target position imposed by non-ideal realistic trap profiles. This trade-off showed the benign effect of typical atom chip anharmonicities on the transport speeds.  For the sake of experimental implementation, the efficiency of this proposal was tested against typical deviations in the main control parameters (magnetic field and timing errors) showing an excellent degree of robustness. Landmark effects of BEC physics as collective excitations were considered and analyzed. The results of this latter investigation revealed the benign character of collective excitations compared to the single particle approach on one hand, and the potential for optimization one could benefit from by using these collective excitations on the other.

Combining all the aforementioned tools, this study exhibits the possibility to precisely transport an atom-chip-generated BEC for several mm with a $\mu$m control level. Delta-kick atom chip collimation would prepare this ensemble in a regime of a pK expansion rate thanks to the collective excitations acquired during the transport ramp. This highly controlled BEC source concept would require only few hundreds of ms, about $200$\,ms for the study case of this article, when implemented in a state-of-the-art atom chip BEC machine. These specifications make of the proposed arrangement an exquisite and novel source concept to feed a highly precise atom interferometer. This would allow to unfold the already promising potential (mobility, autonomy and low power consumption) of atom chip-based atomic sensors in the metrology field \citep{PRLAbend2017}. Further directions would involve the implementation of optimal control theory tools \citep{PRAPeirce1988, PRAHohenester2007} to consider arbitrary potential profiles and even faster manipulations while allowing for larger intermediate excitations. The methods developed in this paper apply directly for optimizing the manipulation of cold atomic ensembles in optical dipole traps. The possibility to generate "painted potentials" \citep{NJPHenderson2009} with these traps is of a particular interest as a future complementary control tool in shortcut-to-adiabaticity protocols. 

%%%%%%%%%%%%%%%%%%%%%%%%%%%%%%%%%%%%%%%%%%%%%%%%%%%%%%%%%%%%%%%%%%%%%%%%
\section*{Acknowledgments}
%%%%%%%%%%%%%%%%%%%%%%%%%%%%%%%%%%%%%%%%%%%%%%%%%%%%%%%%%%%%%%%%%%%%%%%%

This work is supported by the German Space Agency (DLR) with funds provided by the Federal Ministry for Economic Affairs and Energy (BMWi) due to an enactment of the German Bundestag under Grant No. DLR 50WM1552-1557 (QUANTUS-IV- Fallturm). We also acknowledge the use of the computing cluster GMPCS of the LUMAT federation (FR 2764 CNRS). We acknowledge CINES, France for providing access and support to their computing platform Occigen under project AP-010810188. We wish to thank the Collaborative Research Center geo-Q of the Deutsche Forschungsgemeinschaft (SFB 1128), the QUEST-Leibniz-Forschungsschule (QUEST-LFS) and acknowledge financial support from “Niedersächsisches Vorab” through the “Quantum- and Nano-Metrology (QUANOMET)” initiative within the project QT3. RC is grateful to the German Foreign Academic Exchange (DAAD) for partially supporting his research activities in Germany and to the IP@Leibniz program of the Leibniz University of Hanover for travel grants supporting his stays in France. RC and NG acknowledge mobility support from the Q-SENSE project, which has received funding from the European Union's Horizon 2020 Research and Innovation Staff Exchange (RISE) Horizon 2020 program under Grant Agreement Number 691156. Additional mobility funds were thankfully made available through the bilateral exchange project PHC-Procope. We thank Sina Loriani, Jan-Niclas Siem{\ss}, and Tammo Sternke for valuable discussions. NG expresses out appreciation to Christian Schubert for sharing his expertise during the course of this research.

%%%%%%%%%%%%%%%%%%%%%%%%%%%%%%%%%%%%%%%%%%%%%%%%%%%%%%%%%%%%%%%%%%%%%%%%
\appendix
%%%%%%%%%%%%%%%%%%%%%%%%%%%%%%%%%%%%%%%%%%%%%%%%%%%%%%%%%%%%%%%%%%%%%%%%

%%%%%%%%%%%%%%%%%%%%%%%%%%%%%%%%%%%%%%%%%%%%%%%%%%%%%%%%%%%%%%%%%%%%%%%%
\section{Extracting the trap parameters}
\label{AnnexA}
%%%%%%%%%%%%%%%%%%%%%%%%%%%%%%%%%%%%%%%%%%%%%%%%%%%%%%%%%%%%%%%%%%%%%%%%

One can very accurately fit the variation of the trapping frequency with $z_t$ using a second order Pad\'e function in the form
\begin{equation}
\omega_z^2(t)=\frac{\alpha + \beta z_t}{1 + \gamma z_t + \zeta z_t^2}\,.
\label{omega2}
\end{equation}
The classical evolution of the particle is set by Newton's equations, given in Eq.\,(\ref{master_equation_h}). Using Eqs.\,(\ref{omega2})\,and\,(\ref{master_equation_h}), one can infer the evolution of the minimum of the trap $z_t$ as a function of $z_a(t)$ and its derivatives, by solving the simple second order polynomial equation
\begin{equation}
 (\zeta\ddot{z}_a-\beta)z_t^2+(\beta z_a+\gamma\ddot{z}_a(t)-\alpha)z_t+\ddot{z}_a+\alpha z_a=0\,.
\label{poly_order_2}
\end{equation}
The two possible solutions are
\begin{equation}
z_t^{\pm}(t)=\frac{-(\beta z_a(t)+\gamma\ddot{z}_a(t)-\alpha)\pm\sqrt{\Delta(t)}}
                  {2(\zeta\ddot{z}_a(t)-\beta)} 
\label{ztpm}
\end{equation}
where the discriminant $\Delta(t)$ is defined by
\begin{equation}
\Delta(t)=(\beta z_a+\gamma\ddot{z}_a-\alpha)^2-4(\zeta\ddot{z}_a-\beta)(\ddot{z}_a+\alpha z_a)\,.
\end{equation}
One of these two solutions is physically admissible and inserting it in Eq.\,(\ref{omega2}) yields the frequency variation $\omega_z(t)$. Since the bias field $B_{bias}$ can also be easily and accurately fitted by a Pad\'e function of $z_t$, the necessary variation of $B_{bias}(t)$ to perform an STA transport is easily extracted.

%%%%%%%%%%%%%%%%%%%%%%%%%%%%%%%%%%%%%%%%%%%%%%%%%%%%%%%%%%%%%%%%%%%%%%%%
\section{Expansion temperature}
\label{AnnexB}
%%%%%%%%%%%%%%%%%%%%%%%%%%%%%%%%%%%%%%%%%%%%%%%%%%%%%%%%%%%%%%%%%%%%%%%%

In analogy with the common definition of temperature in Maxwell-Boltzmann statistics \citep{PRAGaaloul2006}, we define the expansion temperature by:

\begin{equation}
\frac{3}{2}\,k_B T = \frac{m}{2}\left[ \left(\frac{d\Delta x}{dt}\right)^{\!\!2} + \left(\frac{d\Delta y}{dt}\right)^{\!\!2} + \left(\frac{d\Delta z}{dt}\right)^{\!\!2} \right],
\label{def-T}
\end{equation}
and from Eq.\,(\ref{link_CD-GPE}) we easily obtain
\begin{equation}
k_BT=\frac{m}{21}\left[ \left(\frac{dR_x}{dt}\right)^{\!\!2} + \left(\frac{dR_y}{dt}\right)^{\!\!2} + \left(\frac{dR_z}{dt}\right)^{\!\!2} \right].
\label{TexpBEC}
\end{equation}
Note also that in 1D the coefficient $3/2$ of Eq.\,(\ref{def-T}) is replaced by $1/2$ and we end up in this case with $k_BT=m(dR/dt)^2/7$.

%%%%%%%%%%%%%%%%%%%%%%%%%%%%%%%%%%%%%%%%%%%%%%%%%%%%%%%%%%%%%%%%%%%%%%%%
\bibliographystyle{iopart-num}
\bibliography{NJPFormattest.bib}
%%%%%%%%%%%%%%%%%%%%%%%%%%%%%%%%%%%%%%%%%%%%%%%%%%%%%%%%%%%%%%%%%%%%%%%%

\end{document}